\documentclass[journal]{IEEEtran}
\usepackage{amsmath,amsfonts}
\usepackage{algorithmic}
\usepackage{algorithm}
\usepackage{array}
\usepackage[caption=false,font=normalsize,labelfont=sf,textfont=sf]{subfig}
\usepackage{textcomp}
\usepackage{stfloats}
\usepackage{url}
\usepackage{verbatim}
\usepackage{graphicx}
\usepackage{color}
\usepackage{amssymb}
\usepackage[numbers,sort&compress]{natbib}

\makeatletter

\def\section{\@startsection{section}{1}{\z@}%
{-1.0ex plus -0.2ex minus -0.2ex}%
{0.3ex plus 0.1ex}%
{\normalfont\normalsize\bfseries}}

\def\subsection{\@startsection{subsection}{2}{\z@}%
{-0.8ex plus -0.2ex minus -0.2ex}%
{0.3ex plus 0.1ex}%
{\normalfont\normalsize\itshape}}

\def\subsubsection{\@startsection{subsubsection}{3}{\z@}%
{-0.6ex plus -0.2ex minus -0.2ex}%
{0.2ex plus 0.1ex}%
{\normalfont\normalsize\itshape}}

\makeatother

\AtBeginDocument{
\setlength{\abovedisplayskip}{3pt}
\setlength{\belowdisplayskip}{3pt}
\setlength{\abovedisplayshortskip}{2pt}
\setlength{\belowdisplayshortskip}{2pt}
}
\begin{document}

\title{Phase Uniformity Detector for GRSM Receiver in mmWave and sub-THz bands}

\author{
Oshin Daoud, Haïfa Farès, Amor Nafkha,~\IEEEmembership{Member,~IEEE},\\
Yahia Medjahdi, Laurent Clavier,~\IEEEmembership{Member,~IEEE}
\thanks{O. Daoud, H. Farès, and A. Nafkha are with IETR - UMR CNRS 6164, CentraleSupélec, 35576 Cesson-Sévigné, France (emails: \{oshin.daoud, haifa.fares, amor.nafkha\}@centralesupelec.fr).}

\thanks{Y. Medjahdi and L. Clavier are with IMT Nord Europe, Institut Mines-Télécom, Centre for Digital Systems, F-59653 Villeneuve d’Ascq, France (emails: \{yahia.medjahdi, laurent.clavier\}@imt-nord-europe.fr).}

\thanks{L. Clavier is also with INRIA, Villeneuve d’Ascq, France, and IEMN - UMR CNRS 8520, University of Lille, France.}
}

\markboth{IEEE Wireless Communications Letters}%
{Shell \MakeLowercase{\textit{et al.}}: A Sample Article Using IEEEtran.cls for IEEE Journals}

\IEEEpubid{0000--0000/00\$00.00~\copyright~2026 IEEE}

\maketitle

\begin{abstract}
This paper introduces a phase-domain statistical detector, the Phase Uniformity Detector (PUD), for binary hypothesis testing in Generalized Receive Spatial Modulation (GRSM) systems. The PUD uses direct RF sampling to obtain received signal samples, their phases are modeled via Directional Statistics (DS). A Generalized Likelihood Ratio Test (GLRT) is derived and reduced to a Rayleigh uniformity test with a closed-form, noise-variance-independent threshold. Unlike conventional Energy Detection (ED), the PUD offers robust spatial detection under Independent Local Oscillator Phase Noise (ILO-PN), remaining insensitive to energy fluctuations and noise uncertainty. Additionally, a phase-coherence-aware combining scheme mitigates ILO-PN without requiring estimation.

\end{abstract}


\section{Introduction}
\IEEEPARstart{G}{eneralized} receive spatial modulation (GRSM) has attracted significant attention due to its ability to enhance spectral efficiency while maintaining low receiver hardware complexity and energy consumption. Detection in GRSM involves two stages: identifying the indexed receive antennas (spatial pattern) and detecting the transmitted modulation symbols. Joint maximum-likelihood (ML) detection \cite{b1} is complex; therefore decoupled schemes \cite{b2,b3} emerged to reduce complexity.
In \cite{b4}, a per-antenna amplitude detector was proposed for spatial detection. In \cite{b5}, this was simplified into an energy-sensing approach with a predefined threshold that minimizes the spatial error probability.
To address the stronger Phase Noise (PN) impact at mmWave and sub-THz frequencies, we extended in \cite{b6}, the energy detection framework to different Local Oscillator architectures. Under Independent Local Oscillator (ILO), PN-induced energy fluctuations mismatch the predefined threshold and instantaneous received energy, requiring adaptive threshold updates at higher computational cost.
These approaches assume perfect noise variance knowledge, which is impractical. Noise uncertainty significantly degrades energy detection, especially at low SNR \cite{b7}, motivating the use of signal features beyond energy.
In this context, the received signal phase is a natural angular feature that can be analyzed using Directional Statistics (DS), where each observation is mapped onto the unit circle \cite{b8,b9}. This representation enables hypothesis testing in the angular domain. In \cite{b10,b11}, DS has been successfully applied to blind modulation classification and spectrum sensing.
In GRSM, each receive antenna is modeled under two hypotheses: $\mathcal{H}_0$ (inactive, noise only) and $\mathcal{H}_1$ (active, signal plus noise). In the phase domain, $\mathcal{H}_0$ yields uniformly distributed phases, while $\mathcal{H}_1$ produces phase concentration. This distinction enables DS-based uniformity testing using multiple observations obtained through direct RF sampling.
Building on the above, Energy Detector (ED) in GRSM degrades under ILO-PN and noise uncertainty, and robust low-complexity alternatives remain limited in numbers and performance. Accordingly, we propose a phase-domain statistical detector for GRSM systems. To the best of our knowledge, this is the first work to exploit DS for spatial detection in GRSM. 
The main contributions are summarized as follows:
\textbf{(i)} Phase Uniformity Detector (PUD) is proposed for GRSM detection at low SNR.
\textbf{(ii)} The PUD is derived from a Generalized Likelihood Ratio Test (GLRT) and reduced to a low-complexity Rayleigh uniformity test, exploiting direct RF sampling for hardware implementation.
\textbf{(iii)} A phase-coherence-aware combining scheme exploits a priori information from the proposed PUD to mitigate ILO-PN in MQAM detection without explicit PN estimation.
\textbf{(iv)} Spatial and MQAM BER results under No-PN and ILO-PN demonstrate the superiority of the PUD over ED at low SNR and the effectiveness of the proposed combining scheme in mitigating ILO-PN.

\IEEEpubidadjcol
\section{System Model}
\subsection{The Transmitter}
The transmitter, as given in our previous works \cite{b5,b6}, is a fully digital Base Station (BS) equipped with $N_t$ antennas and  $N_t$ RF chains. The incoming binary stream is partitioned into two parts: the first $N_a$ bits represent the spatial pattern, while the remaining $\log_2(M)$ bits encode the MQAM-modulated symbol. The resulting Spectral Efficiency (SE) is  
\(
\text{SE} = N_a + \log_2(M) \quad \text{bits/s/Hz}.
\)
In this work, $\text{SE}=8$ bits/s/Hz, the number of spatial bits is $N_a=4$ for $M = 16$.
In the Time Division Duplexing (TDD) downlink, the transmitter estimates the Channel State Information (CSI) by exploiting channel reciprocity. Assuming perfect CSI at the BS, a Receiver antenna Selection (RAS) algorithm is employed to select the $N_a$ least correlated receive antennas. As a result, the channel matrix \( \mathbf{H} \in \mathbb{C}^{N_r \times N_t} \) is reduced to \( \mathbf{H}_a \in \mathbb{C}^{N_a \times N_t} \).
 The $N_a$ bits of each spatial symbol are then mapped to the selected receive antennas using a Zero-Forcing (ZF) precoder \( \mathbf{B} \in \mathbb{C}^{N_t \times N_a} \). 
The transmitter PN is noted as in \cite{b6}:
\begin{equation}
  \boldsymbol{\Phi}_{\text{tx}} = \operatorname{diag}\left( \left[e^{j\phi_1}, \cdots, e^{j\phi_{N_t}} \right]^{\mathrm{T}} \right),
  \label{eq:1}
\end{equation}
where $\phi_l \sim \mathcal{N}(0,\sigma_{\mathrm{pn}}^2)$, represents the independent PN sample at the $l-{th}$ transmit antenna. Under the ILO architecture, the PN samples $\{\phi_l\}_{l=1}^{N_t}$ are statistically independent.
The channel follows an extended Saleh-Valenzuela model, capturing the propagation environment as a clustered channel composed of Line-of-Sight (LOS) and Non-LOS components \cite{b12}.
\subsection{The Receiver}
The received signal during the symbol duration $T_{sym}$ in ideal (No-PN) case:
\begin{equation}
\begin{aligned}
\mathbf{y}(t)
&= \sqrt{\alpha}\,\mathbf{H}_a\mathbf{B}\mathbf{s}_i x_j+\mathbf{w}(t),\\[-3pt]
y_k(t)
&= \sqrt{\alpha}\,|x_j|e^{j\theta_x}s_{ik}+w_k(t).
\end{aligned}
\label{eq:2}
\end{equation}
where $\mathbf{y}(t) \in \mathbb{C}^{N_a \times 1}$ denotes the received signal vector at time instant $t$ within the symbol duration $T_{\text{sym}}$, and $\mathbf{w}(t) \sim \mathcal{CN}(\mathbf{0}, \sigma^2 \mathbf{I})$ is additive white Gaussian noise (AWGN) with variance $\sigma^2$. 
The normalization factor is given by
\(
\alpha = \mathbb{E}_{\mathbf H_a}
\left\{
\left[\operatorname{tr}\left((\mathbf H_a \mathbf H_a^{H})^{-1}\right)\right]^{-1}
\right\}.
\)
The transmitted symbol $x_j$ belongs to the $M$-QAM constellation
$\mathcal{C} = \{x_j \mid j = 1,\dots,M\}$ with amplitude $|x_j|$ and phase $\theta_x$.
The spatial pattern $\mathbf{s}_i \in \{0,1\}^{N_a \times 1}$ defines the antenna activation pattern, where $i \in \{1,\dots,2^{N_a}-1\}$.
Finally, $y_k(t)$ denotes the received signal at the $k$-th antenna, $s_{ik}$ is the $k$-th bit of the spatial pattern, and $w_k(t)$ is AWGN at the $k$-th receive antenna.
Under ILO-PN, the received signal becomes:
\begin{equation}
\begin{aligned}
\mathbf{y}(t) &= \sqrt{\alpha}\,\mathbf{H}_a \boldsymbol{\Phi}_{\text{tx}} \mathbf{B} \, s_i x_j + \mathbf{w}(t), \\[-3pt]
y_k(t) &= \sqrt{\alpha}\, |x_j| e^{j\theta_x}
\sum_{l=1}^{N_t} H_a(k,l)\, e^{j\phi_l}\, B(l,k)\, s_{ik} + w_k(t).
\end{aligned}
\label{eq:3}
\end{equation}
The term $H_a(k,l)$ denotes the channel coefficient between the $l$-th transmit antenna and the $k$-th receive antenna, while $B(l,k)$ represents the ZF precoding matrix element. 
During $T_{\text{sym}}$, the channel, transmitted symbol in  \eqref{eq:2} and \eqref{eq:3} and ILO-PN in  \eqref{eq:3}, are assumed quasi-static. The latter assumption is adopted in mmWave and sub-THz communication systems, where the large available bandwidths (typically hundreds of MHz to several GHz) lead to symbol durations in the nanosecond order. Since these symbol durations are much shorter than the time scale over which the oscillator phase evolves, the PN process can be considered approximately constant within a symbol interval.
It is worth mentioning that the received signal $\mathbf{y}$ in \eqref{eq:2}, \eqref{eq:3}, is considered in the baseband. The corresponding RF signal at the carrier frequency $f_c$ can be expressed as \cite{b13}:
\begin{equation}
\mathbf{y}^{\mathrm{RF}}(t) = \sqrt{2}\,\Re  \left\{ \mathbf{y}^{\mathrm{BB}}(t)\, e^{j2\pi f_c t} \right\}.
\label{eq:4}
\end{equation}
By sampling the RF signal at a rate $F_s$, the discrete-time representation becomes
\begin{equation}
\mathbf{y}_n^{\mathrm{RF}} = \sqrt{2}\,\Re \left\{ \mathbf{y}_n^{\mathrm{BB}}\, e^{j2\pi (f_c/F_s)n} \right\}.
\label{eq:5}
\end{equation}
The receiver operates in two stages. Spatial detection is performed independently at each antenna using binary hypothesis testing. Sec.~\ref{sec:III} presents the baseline $N$-sample energy detector (N-ED), and Sec.~\ref{sec:IV} introduces the proposed Phase Uniformity Detector (PUD). The detected active branches are combined and the transmitted MQAM symbol is recovered using ML detection, as in Sec.~\ref{sec:V}.

\section{N-Samples Energy Detector (N-ED)}
\label{sec:III}
\begin{figure}[t]
    \centering
    \includegraphics[width=\columnwidth]{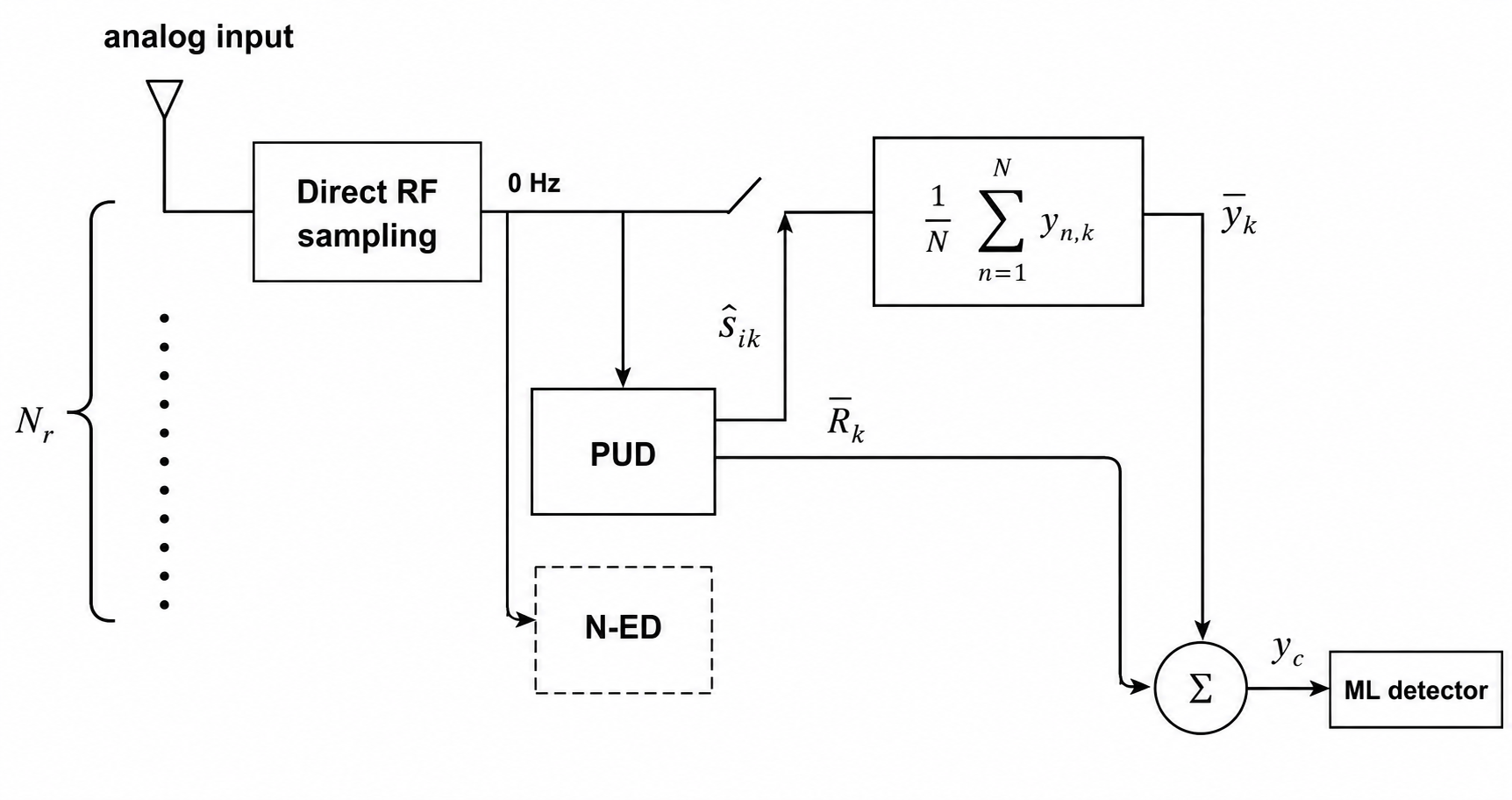}
    \caption{The proposed receiver branch structure based on Direct RF Sampling.}
    \label{fig:1}
\end{figure}

In this work, an $N$-sample energy detector (N-ED) collects $N$ temporal samples per receive antenna over one symbol duration $T_{\text{sym}}$, with energy computed in either the analog or digital domain \cite{b14}, Figure ~\ref{fig:1}.
Unlike conventional N-ED receivers that rely on mixer-based downconversion, the proposed architecture adopts direct RF sampling, where the received signal is sampled directly at RF and digitally translated to an intermediate frequency using a numerically controlled oscillator (NCO). Recent advances in high-speed ADC technologies have made direct RF sampling increasingly feasible for mmWave and sub-THz systems. In particular, time-interleaved ADC architectures have emerged as a promising solution for achieving the high sampling rates required at these frequencies \cite{b15}.

\subsection{Mathematical Model}
This detector is modeled and the detection threshold is designed based on the ideal No-PN case. Recalling the $k_{th}$ received signal in \eqref{eq:2} and applying \eqref{eq:4}, the $k_{th}$ RF received signal is given:
\[
y_k^{\mathrm{RF}}(t)
= \sqrt{2}\,\Re \left\{
\left(
\sqrt{\alpha}\, |x_j| e^{j\theta_x} s_{ik} + w_k(t)
\right)
e^{j2\pi f_c t}
\right\}.
\]
Under $\mathcal{H}_0$ ($s_{ik}=0$), the RF signal is Gaussian with zero mean:
\(
\sqrt{2}\, y_k^{\mathrm{RF}}(t) \sim \mathcal{N}(0,\sigma^2).
\)
Under $\mathcal{H}_1$ ($s_{ik}=1$), the RF signal remains Gaussian with a non-zero mean:
\(
\sqrt{2}\, y_k^{\mathrm{RF}}(t) \sim \mathcal{N}(\mu(t), \sigma^2),
\)
where
\(
\mu(t) = \sqrt{\alpha}\, \Re\{x_j e^{j2\pi f_c t}\}.
\)
After direct RF sampling and digital downconversion to baseband, the discrete-time samples are given by:
\begin{equation}
{y}_k^{\mathrm{BB}} = \left[ y_k^{\mathrm{BB}}(1), y_k^{\mathrm{BB}}(2), \dots, y_k^{\mathrm{BB}}(N) \right]
\label{eq:6}
\end{equation}
Under $\mathcal{H}_0$, the samples follow:
\(
y_k(n) \sim \mathcal{CN}(0,\sigma^2),
\)
while under $\mathcal{H}_1$:
\(
y_k(n) \sim \mathcal{CN}(\mu_k,\sigma^2),
\)
where $\mu_k = \sqrt{\alpha}\, x_j$.
The normalized energy measured by the $N$-sample energy detector (N-ED) at the $k$-th branch is given by:
\(
E_k = \frac{1}{\sigma^2} \sum_{n=1}^{N} |y_k(n)|^2.
\)
Under $\mathcal{H}_0$, $E_k$ follows a central chi-square distribution, while under $\mathcal{H}_1$, it follows a non-central chi-square distribution, both with $2N$ degrees of freedom.

\subsection{Threshold Calculation}
For a target false alarm probability $P_{fa} = \Pr(E_k > \gamma \mid \mathcal{H}_0)$, the threshold $\gamma$ is determined from the  Complementary Cumulative Distribution Function (CCDF):
\(
P_{fa} = 1 - F_{\chi_{2N}^2}(\gamma),
\)
where $F_{\chi_{2N}^2}(\cdot)$ is the CDF of the central chi-square distribution with $2N$ degrees of freedom.
Solving for $\gamma$ yields:
\begin{equation}
\gamma = F_{\chi_{2N}^2}^{-1}(1 - P_{fa}),
\label{eq:7}
\end{equation}
where $F_{\chi_{2N}^2}^{-1}(\cdot)$ is the inverse CDF of the central chi-square distribution.
Finally, the spatial bit detection rule based on the N-ED is given by: $\hat{s}_{ik}=1$ (i.e., $\mathcal{H}_1$) if $E_k > \gamma$, and $\hat{s}_{ik}=0$ (i.e., $\mathcal{H}_0$) otherwise,
where $\hat{s}_{ik}$ is the estimated spatial bit at the $k$-th receive antenna. The detection is performed independently across the $N_a$ receive branches, and the estimated spatial pattern is reconstructed as $\hat{\mathbf{s}}_i = [\hat{s}_{i1},.., \hat{s}_{ik},.., \hat{s}_{iN_a}]^T$.
Measuring $E_k$ with N-ED, assumes perfect knowledge of the noise variance at the receiver, which is not the case in practical implementations, especially in low-SNR regimes. This uncertainty shifts the decision threshold thereby deteriorating detection accuracy.
To model this imperfection, we adopt the bounded noise uncertainty model \cite{b7}, where the estimated noise variance $\widehat{\sigma}^2$ lies within
\(
\widehat{\sigma}^2 \in
\left[
\frac{1}{1+\rho}\sigma^2,\;
(1+\rho)\sigma^2
\right],
\)
where $\sigma^2$ is the true noise variance and $\rho \ge 0$ is the noise uncertainty parameter. The corresponding uncertainty level in dB is
\(
\rho_{\mathrm{dB}}
=
10\log_{10}(1+\rho).
\)

\section{The Proposed Spatial Detector:\\ Phase Uniformity Detector (PUD)}
\label{sec:IV}
The objective of the proposed Phase Uniformity Detector (PUD) is to distinguish between $\mathcal{H}_0$ and $\mathcal{H}_1$ by exploiting the statistical characteristics of the sampled received-signal phases, without requiring knowledge of the noise variance.
The detection rule is therefore based on the statistical behavior of the phase samples obtained from \eqref{eq:6}:
\begin{equation}
\theta_k(n) = \angle y_k^{\mathrm{BB}}(n), \quad n = 1,2,\dots,N.
\label{eq:8}
\end{equation}
where $\angle(\cdot)$ denotes the phase operator. Thus, $\theta_k(n)$ represents the phase of the $n_{th}$ baseband signal sample on the $k_{th}$ branch.
Based on \eqref{eq:2} and \eqref{eq:3}, specifically, under $\mathcal{H}_0$ ($s_{ik}=0$), the $n_{th}$ sample is: $y_k(n) = w_k(n)$. 
In contrast, under $\mathcal{H}_1$ ($s_{ik}=1$), the $n_{th}$ sample is $y_k(n) = z_k(n) + w_k(n)$, where $z_k$ represents the signal component in two cases: $z_k$ is expressed as $z_{k,\mathrm{No\text{-}PN}} = \sqrt{\alpha}\, x_j$, and $z_{k,\mathrm{ILO}} = \sqrt{\alpha}\, x_j
\sum_{l=1}^{N_t} H_a(k,l)\, e^{j\phi_l}\, B(l,k)\,$. 
Accordingly, the phase samples defined in \eqref{eq:8} can be written:
\begin{equation}
\theta_k(n) =
\begin{cases}
\angle w_k(n), & \mathcal{H}_0, \\
\angle \big( z_k(n) + w_k(n) \big), & \mathcal{H}_1.
\end{cases}
\label{eq:9}
\end{equation}
The fundamental distinction between $\mathcal{H}_0$ and $\mathcal{H}_1$, is starting from defining the statistical behaviours of phases in \eqref{eq:9} based on directional statistics. 
This is detailed in the next subsection.
\subsection{Phase Statistics}
\begin{figure}[t]
    \centering
    \includegraphics[width=0.8\columnwidth]{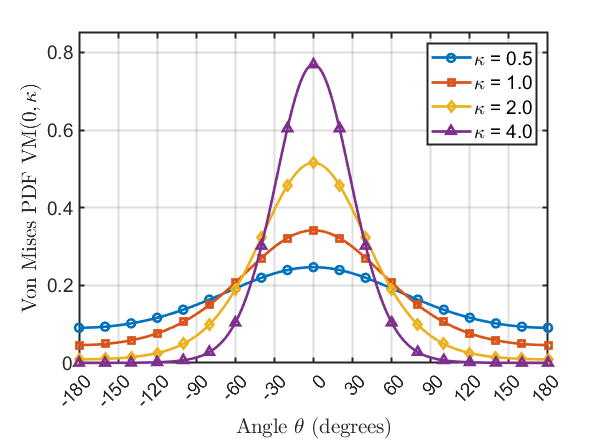}
    \caption{The PDF of Von Mises distribution, $\kappa = 0.5,1,2,4$ \cite{b8}.}
    \label{fig:vonmises}
\end{figure}

Under $\mathcal{H}_0$ (only noise), considering the phases as in \eqref{eq:9}, they are uniformly distributed over $[-\pi,\pi)$. Hence, the phase samples are i.i.d. with probability density function (PDF) \cite{b8}:
\begin{equation}
f_{\theta \mid \mathcal{H}_0}(\theta) = \frac{1}{2\pi}, \quad \theta \in [-\pi,\pi).
\label{eq:10}
\end{equation}
Under $\mathcal{H}_1$ (signal and noise), the presence of $z_k$ introduces a structured phase component. For the configurations (No-PN, ILO-PN), this results in directional phase behavior, i.e. the phases as in \eqref{eq:9} are clustered around a mean direction despite possible rotations or spreading induced by PN. Following standard DS practice, this clustering is approximated by a Von Mises distribution, which provides an accurate and analytically tractable model for phase concentration \cite{b8}, with PDF:
{\small
\begin{equation}
f_{\theta \mid \mathcal{H}_1}(\theta;\mu,\kappa) = \frac{1}{2\pi I_0(\kappa)} \exp\big( \kappa \cos(\theta - \mu) \big), \quad \theta \in [-\pi,\pi),
\label{eq:11}
\end{equation}}
where $\mu$ is the mean direction, $\kappa$ is the concentration parameter, and $I_0(\kappa)$ is the modified Bessel function of order zero. It is worth noting that the Von Mises distribution models phase samples clustered around a preferred direction when the concentration parameter $\kappa > 0$, as depicted in Figure \ref{fig:vonmises}, whereas it reduces to the uniform distribution on the circle when $\kappa = 0$~\cite{b8,b9}.

\subsection{Generalized Likelihood Ratio Test (GLRT)}

After characterizing the phase distributions under both hypotheses, it is observed that $\mathcal{H}_1$ is a composite hypothesis since the parameters of the Von Mises distribution are unknown; i.e, the mean direction $\mu$ and concentration parameter $\kappa$. Therefore, the classical likelihood ratio test (LRT) can not be directly applied, and GLRT is adopted for decision-making in hypothesis testing \cite{b16,b17,b18}. The GLRT compares the maximum likelihood under $\mathcal{H}_1$ with the likelihood under $\mathcal{H}_0$ as:
{
\setlength{\abovedisplayskip}{1pt}
\setlength{\abovedisplayshortskip}{1pt}
\setlength{\belowdisplayskip}{3pt}
\setlength{\belowdisplayshortskip}{3pt}
\begin{equation}
\Lambda_{\mathrm{GLR}}(\boldsymbol{\theta})
=
\frac{\displaystyle \max_{\mu,\,\kappa} L_1(\boldsymbol{\theta}; \mu, \kappa)}
{L_0(\boldsymbol{\theta})}
\mathop{\gtrless}\limits_{\mathcal{H}_0}^{\mathcal{H}_1}
\gamma ,
\label{eq:12}
\end{equation}
}
where $L_1(\boldsymbol{\theta};\mu,\kappa)$ and $L_0(\boldsymbol{\theta})$ denote the likelihood functions under $\mathcal{H}_1$ and $\mathcal{H}_0$, respectively, and $\gamma$ is the decision threshold.
For $N$ independent phases, from \eqref{eq:10}, the likelihood under $\mathcal{H}_0$:
\begin{equation}
L_0(\boldsymbol{\theta}) = \prod_{n=1}^N \frac{1}{2\pi} = \left( \frac{1}{2\pi} \right)^N,
\label{eq:13}
\end{equation}
while the likelihood under $\mathcal{H}_1$, using \eqref{eq:11}, is given by:
{
\setlength{\abovedisplayskip}{1pt}
\setlength{\abovedisplayshortskip}{1pt}
\setlength{\belowdisplayskip}{3pt}
\setlength{\belowdisplayshortskip}{3pt}
\begin{equation}
\begin{aligned}
L_1(\boldsymbol{\theta}; \mu, \kappa)
&= \prod_{n=1}^N \frac{1}{2\pi I_0(\kappa)} \exp\big(\kappa \cos(\theta_n - \mu)\big). 
\end{aligned}
\label{eq:14}
\end{equation}}
Substituting \eqref{eq:13} and \eqref{eq:14} into \eqref{eq:12} and taking the natural logarithm, the log-GLRT:
{\small
\begin{equation}
\ln \Lambda 
=
\max_{\mu,\kappa} \left[
- N \ln I_0(\kappa)
+ \kappa \sum_{n=1}^N \cos(\theta_n - \mu)
\right]\mathop{\gtrless}\limits_{\mathcal{H}_0}^{\mathcal{H}_1}
\ln \gamma.
\label{eq:15}
\end{equation}}
By maximizing the log-likelihood under $H_1$ with respect to the unknown $\mu$ and $\kappa$, the GLRT transforms the composite hypothesis into a simple hypothesis parameterized by the maximum-likelihood estimates (MLEs) of $\mu$ and $\kappa$\cite{b18}.
Following the standard MLEs procedure for the von Mises distribution in \cite[Sec.~5.3.1]{b8}, the likelihood is first maximized with respect to $\mu$, and subsequently with respect to $\kappa$.
\subsection{Maximum Likelihood Estimations (MLEs)}
\begin{itemize}
\item {Maximization over $\mu$} \cite{b15}:  
From \eqref{eq:15}, taking the left term of the inequality, for a fixed $\kappa$, the term to be maximized is $\sum_{n=1}^N \cos(\theta_n - \mu)$. Defining the resultant vector
\(
\mathbf{R} = \sum_{n=1}^{N} e^{j\theta_n}\), its length  \( R = |\mathbf{R}|\). This implies that:
{\small
\[
\sum_{n=1}^{N} \cos(\theta_n - \mu)
= \Re\{ e^{-j\mu} \mathbf{R} \}
= R \cos(\angle \mathbf{R} - \mu).
\]}
The maximum is achieved for
\(
\hat{\mu} = \angle \mathbf{R}
= \arg \left( \sum_{n=1}^{N} e^{j\theta_n} \right),
\) yields $\max_{\mu} \sum_{n=1}^{N} \cos(\theta_n - \mu) = R$.
Substituting into \eqref{eq:15}, the log-GLR becomes:
\begin{equation}
\ln \Lambda
= \max_{\kappa} \left[ - N \ln I_0(\kappa) + \kappa R \right],
\label{eq:16}
\end{equation}
where \( R = \left| \sum_{n=1}^{N} e^{j\theta_n} \right| \) is the resultant vector length and its mean:
\(
\bar{R}=\frac{R}{N},
\) which measures the concentration of the phase samples around the mean direction.
The mean resultant length satisfies $0 \leq \bar{R} \leq 1$. As $\bar{R}\rightarrow 1$, the phase samples become highly concentrated around a common direction, indicating a strong deviation from uniformity. Conversely, as $\bar{R}\rightarrow 0$, the phase samples approach a uniform distribution on the unit circle.
\item{Maximization over $\kappa$} \cite{b18}: 
As stated in \eqref{eq:16}, after substituting the MLE of $\mu$, the log-GLR reduces to a function of $\kappa$ only:
\[
f(\kappa)=\kappa R - N\ln I_0(\kappa), \quad \kappa \ge 0.
\]
The optimal value of $\kappa$ is obtained by maximizing $f(\kappa)$. Taking the first derivative and setting it to zero yields the necessary condition
\(
f'(\kappa)= R - N \frac{I_1(\kappa)}{I_0(\kappa)} = 0\),
which leads to
\(
\frac{I_1(\kappa)}{I_0(\kappa)}=\frac{R}{N}
\). Assuming $A(\kappa)= \frac{I_1(\kappa)}{I_0(\kappa)}$, the unique maximum-likelihood estimate is
\(
\hat{\kappa}=A^{-1}\!\left(\frac{R}{N}\right).
\)\footnotemark[1] \footnotemark[2]
\footnotetext[1]{
To verify that this stationary point corresponds to a unique global maximum, we examine the concavity of $f(\kappa)$ via its second derivative:
\(
f''(\kappa)=\frac{d}{d\kappa}f'(\kappa)
= -N\frac{d}{d\kappa}\left(\frac{I_1(\kappa)}{I_0(\kappa)}\right).
\)
 In \cite[(Eq.~3.5.32)]{b8}, $A(\kappa)$ is continuous and strictly increasing for $\kappa \ge 0$, hence $A'(\kappa)>0$. Therefore,
\(
f''(\kappa) = -N A'(\kappa) < 0, \quad \forall \kappa \ge 0,
\)
which shows that $f(\kappa)$ is strictly concave on $[0,\infty)$. This guarantees that the solution of $f'(\kappa)=0$ is the unique global maximizer.
}
\footnotetext[2]{
As the function $A(\kappa)$ is continuous and strictly increasing on $[0,\infty)$, with $A(0)=0$ and $\lim_{\kappa\to\infty} A(\kappa)=1$. Hence, $A(\kappa)$ defines a bijection from $[0,\infty)$ onto $[0,1)$, and therefore admits a unique inverse, \cite{b8}, Appendix~1, Eqs.~(A11)–(A14). 
}
\end{itemize}

\subsection{Reduction of the GLRT to the Rayleigh Test}

Substituting the MLE $\hat{\kappa}$ into \eqref{eq:16}
yields:
\begin{equation}
\ln \Lambda
=
\hat{\kappa} R
-
N\ln I_0(\hat{\kappa})
\label{eq:17}
\end{equation}
Equation~\eqref{eq:17} shows that the log-GLR ($\ln \Lambda$) statistic only depends on the resultant length $R$. Because $A(\kappa)$ is strictly increasing\footnotemark[2], $\hat{\kappa}$ is a monotonic function of $R$; hence, the testing of log-GLR in ~\eqref{eq:17}, used to distinguish between $\mathcal{H}_0$ and $\mathcal{H}_1$, is equivalent to a threshold test on $R$, or equivalently on the mean resultant length $\bar{R}=R/N$.
Since $\bar{R}$ quantifies the concentration of the phase samples around their mean direction, the hypothesis test can be interpreted as a test for phase concentration. From a directional statistics perspective, this is equivalent to testing whether the phase samples are uniformly distributed on the unit circle. The Rayleigh test is the classical uniformity test based on the mean resultant length $\bar{R}$; therefore, the GLRT is equivalent to a Rayleigh test~\cite{b8}.
\subsection{Rayleigh Test and threshold design}
The Rayleigh test statistic as in~\cite{b8} is \( T = 2N\bar{R}^{2} = \frac{2}{N}(C^{2}+S^{2}) \),
where $C=\sum_{n=1}^N \cos\theta_n$, $S=\sum_{n=1}^N \sin\theta_n$.
Under $\mathcal{H}_0$, the phases are i.i.d. uniform over $(-\pi,\pi)$, yielding $\mathbb{E}[\cos\theta_n]=\mathbb{E}[\sin\theta_n]=0$ and $\mathrm{Var}(\cos\theta_n)=\mathrm{Var}(\sin\theta_n)=1/2$. For enough $N$, by the Central Limit Theorem, $C$ and $S$ are independent Gaussian random variables with variance $N/2$, leading to
\(
T \sim \chi^2_2 \quad \text{under } \mathcal{H}_0.
\)
Hence, the false-alarm probability is:
\(
P_{\mathrm{FA}} = \Pr(T>\gamma|\mathcal{H}_0)=e^{-\gamma/2},
\)
which yields the threshold
\(
\gamma = -2\ln P_{\mathrm{FA}}.
\)
The Rayleigh test finally is expressed as: 
\begin{equation}
\small
2N\bar{R}^2 \;\gtrless_{\mathcal{H}_0}^{\mathcal{H}_1}\; -2\ln P_{\mathrm{FA}}
\;\;\Longrightarrow\;\;
\bar{R} \;\gtrless_{\mathcal{H}_0}^{\mathcal{H}_1}\; \sqrt{\frac{-\ln P_{\mathrm{FA}}}{N}}.
\label{eq:18}
\end{equation}
It is obvious that the threshold is fixed and independent of the noise power. This significantly reduces the receiver complexity while enhancing robustness to noise uncertainty.

\section{Combining Schemes and MQAM Detection}
\label{sec:V}
After spatial detection, the signals received on the estimated active branches are combined prior to MQAM detection. In Figure \ref{fig:1}, each branch is represented by the sample average of the corresponding baseband observations. For the $k$-th active branch, this average is defined as
\(
\bar{y}_k
=
\frac{1}{N}
\sum_{n=1}^{N}
y_{nk}
\).
\textbf{Classical Combining Scheme}
Following our previous works~\cite{b5,b6}, this approach assumes that all detected active branches contribute equally to the final MQAM decision. Let the set of detected active branches as
\(
\mathcal{A}=\{\,k:\hat{s}_{ik}=1\,\}.
\)
The combined signal is then obtained by combining the averaged samples of all estimated active branches:
\( y_c = \frac{\sum_{k \in \mathcal{A}} \bar{y}_k}{|\mathcal{A}|}. \)
\textbf{Phase-Coherence-Aware Combining}
The proposed combining scheme is fed by a priori information already computed by the PUD. This information is the mean resultant length $\bar{R}$ of the phase samples on the $k$-th active branch expressed as:
\(
\bar{R}_k = \frac{R_k}{N}.
\)
Using the sum of $\bar{R}_k$ over branches as a normalization factor:
{
\setlength{\abovedisplayskip}{6pt}
\setlength{\belowdisplayskip}{6pt}
\setlength{\abovedisplayshortskip}{6pt}
\setlength{\belowdisplayshortskip}{6pt}
\begin{equation}
y_c =
\smash{\frac{\sum_{k \in \mathcal{A}} \bar{y}_k}
{\sum_{k \in \mathcal{A}} \bar{R}_k}}.
\label{eq:19}
\end{equation}}
To analytically justify the proposed combining approach, a decomposition of the average per branch $\bar{y}_k$ is done:
{
\setlength{\abovedisplayskip}{8pt}
\setlength{\belowdisplayskip}{8pt}
\setlength{\abovedisplayshortskip}{8pt}
\setlength{\belowdisplayshortskip}{8pt}
\begin{equation}
\bar{y}_k =
\smash{\frac{1}{N}\sum_{n=1}^{N} A_{nk} e^{j\theta_{nk}}}
\approx
A_k
\smash{\frac{1}{N}\sum_{n=1}^{N} e^{j\theta_{nk}}},
\label{eq:20}
\end{equation}}
where $A_{nk}$ and $\theta_{nk}$ denote the amplitude and phase of the $n$th sample on the $k$th branch, respectively. Since the amplitude is assumed to remain approximately constant over the observation interval, the sample index $n$ is omitted, yielding $A_{kn} \approx A_k$.
Using the DS representation considering $\mu_k$ the mean direction of samples phases on $k_{th}$ branch:
\(
\smash{\frac{1}{N}
\sum_{n=1}^{N}
e^{j\theta_{nk}}}
=
\bar{R}_k e^{j\mu_k}.
\)
The branch average in ~\eqref{eq:20} can be expressed as:
\begin{equation}
\bar{y}_k
\approx
A_k \bar{R}_k e^{j\mu_k},
\label{eq:21}
\end{equation}
implying
\(
|\bar{y}_k|
\approx
A_k \bar{R}_k,
\)
which shows that the magnitude of the averaged signal is directly proportional to the phase coherence of the branch.
Substituting \eqref{eq:21} into \eqref{eq:19} yields
\begin{equation}
y_c
\approx
\frac{\sum_{k\in\mathcal A}
A_k \bar R_k e^{j\mu_k}}
{\sum_{k\in\mathcal A}\bar R_k}.
\label{eq:22}
\end{equation}
In the ideal case, where all branches exhibit perfect phase coherence ($\bar R_k=1$, $\forall k\in\mathcal A$), it reduces to $|\mathcal A|$, corresponding to the number of detected active branches. In the presence of phase dispersion due to PN effect, the contribution of the $k$th branch can be expressed through the coefficient
\(
\frac{\bar R_k}
{\sum_{k\in\mathcal A}\bar R_k}.
\)
Consequently, branches with larger $\bar R_k$ (stronger phase concentration) contribute proportionally more to the combined signal, while branches with lower phase coherence are naturally assigned smaller weights. Therefore, the normalization in \eqref{eq:19} converts the coherence metrics into relative combining weights, ensuring that the resulting signal is a properly scaled coherence-weighted average of the underlying branch signals.
After combining, the transmitted MQAM symbol is detected using the ML rule:
\begin{equation}
\hat{x}_j
=
\arg\min_{x_j \in \mathcal{C}}
\left| y_c - x_j \right|^2.
\label{eq:23}
\end{equation}

\section{Implementation and Results }

\begin{figure}[t]
    \centering
    \includegraphics[width=0.8\columnwidth]{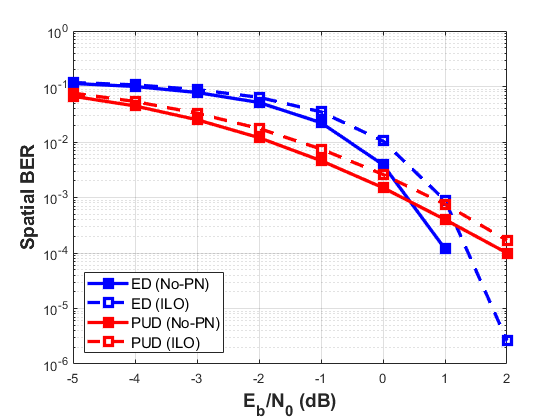}
  \caption{Spatial BER comparison between the proposed PUD and N-ED under No-PN and ILO-PN conditions. The N-ED is evaluated under a bounded noise-uncertainty model. }
    \label{fig:2}
\end{figure}

\begin{figure}[t]
    \centering
    \includegraphics[width=0.8\columnwidth]{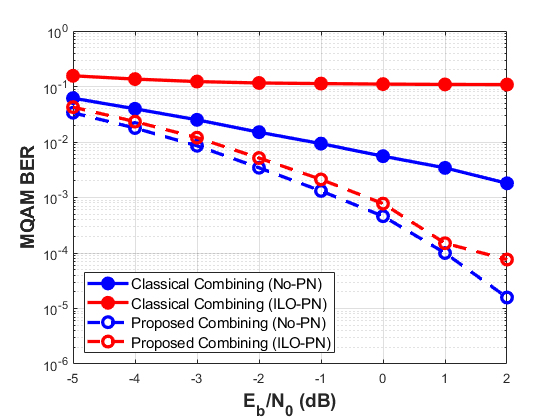}
\caption{MQAM BER of the proposed PUD using the classical/ proposed combining schemes under No-PN and ILO-PN conditions.}
    \label{fig:3}
\end{figure}

The spatial detection using PUD is performed using $N=32$ samples per symbol and a target false-alarm probability of $P_{\mathrm{FA}}=10^{-4}$, yielding a Rayleigh-test threshold of $\gamma=0.536$. For comparison, the N-ED is evaluated under the same parameters, considering a bounded noise-uncertainty model with $\rho_{\mathrm{dB}}=2$ dB.
To evaluate the robustness of the proposed detector under severe PN conditions, the simulations adopt a relatively strong PN variance of $\sigma_{\mathrm{pn}}^{2}=0.1$ rad$^{2}$ as in \cite{b19}.
The results compare the spatial BER of the proposed PUD and N-ED (Fig.~\ref{fig:2}), and the MQAM BER of the proposed and conventional combining schemes (Fig.~\ref{fig:3}).
Figure~\ref{fig:2} highlights the robustness of the proposed PUD for spatial detection. Compared with the practical N-ED under bounded noise uncertainty, the PUD achieves lower BER, particularly at low SNR. Moreover, under ILO-PN, its performance remains close to the No-PN case since it exploits phase concentration rather than instantaneous energy. At higher SNR, however, the N-ED achieves lower BER. This crossover behavior motivates a hybrid strategy that employs the PUD at low SNR and the N-ED at high SNR.
Figure~\ref{fig:3} highlights the impact of the combining strategy on MQAM BER. With conventional equal-gain combining, a pronounced error floor appears under strong ILO-PN due to phase dispersion among the combined replicas. In contrast, the proposed phase-coherence-aware combining suppresses the ILO-PN effect, allowing the MQAM BER to approach the No-PN performance and demonstrating the potential of DS for low-complexity PN mitigation without explicit PN estimation.

\section{Conclusion}
Based on DS, a novel spatial detector (PUD) for GRSM is proposed, providing robust detection under ILO-PN in the low-SNR regime. In addition, a phase-coherence-aware combining scheme is introduced for MQAM detection, mitigating ILO-PN without explicit PN estimation.

\section{acknowledgement: }
This work was supported in part by the Agence Nationale de la Recherche through 'ANR-PEPR Networks of the Future' under Grant NF-SYSTERA 22-PEFT-0006.

\vspace{-11mm}

\vspace{8ex}   

\end{document}